\documentclass[12pt]{article}
\usepackage{mathrsfs}
\usepackage{amsmath,amssymb}

\usepackage{amsfonts}
\usepackage{latexsym}
\usepackage{amsthm}
\usepackage{pictex}

\usepackage[T2A]{fontenc}     % ‚ÌÛÚÂÌÌˇˇ {{T2A}} ÍÓ‰ËÓ‚Í‡ {{TeX}}
\usepackage[cp1251]{inputenc} % ÍÓ‰ËÓ‚Í‡ - ÏÓÊÌÓ ËÒÔÓÎ¸ÁÓ‚‡Ú¸ {{[cp866] [koi8-r]}}
\usepackage{graphicx}

\newcommand{\cz}{{\mathbb C}}
\newcommand{\nz}{{\mathbb N}}

\newcommand{\rz}{{\mathbb R}}
\newcommand{\zz}{{\mathbb Z}}

\def\fourior{Fourier integral operator{}}

\def\neigh{neighborhood{}}

\def\pseudor{pseudodifferential operator{}}

\def\wrt{with respect to}

\def\Re{{\rm Re\,}}
\def\Im{{\rm Im\,}}
\def\lb{\label}
\def\dSRN{de Sitter-Reissner-Nordstr{\"o}m }
\def\dSS{de Sitter-Schwarzschild }
\newtheorem{theorem}{Theorem}

\newtheorem{lemma}{Lemma}

\newtheorem{proposition}{Proposition}

\def\a{\alpha}

\def\G{\Gamma} \def\cD{{\cal D}}

  \def\cH{{\cal H}}   
   \def\cI{{\cal I}}   
\def\p{\psi}

\def\n{\nu}    \def\cP{{\cal P}}   
      
\def\s{\sigma} \def\cR{{\cal R}}   
\def\S{\Sigma}

\def\vp{\varphi}

\def\R{{\Bbb R}}
\def\C{{\Bbb C}}

\def\N{{\Bbb N}}
\def\S{{\Bbb S}}

\def\qq{\quad}
\newcommand{\ma}{\begin{pmatrix}}
\newcommand{\am}{\end{pmatrix}}
\newcommand{\ca}{\begin{cases}}
\newcommand{\ac}{\end{cases}}

\let\geq\geqslant
\let\leq\leqslant
\def\ma{\left(\begin{array}{cc}}
\def\am{\end{array}\right)}

\let\geq\geqslant
\let\leq\leqslant
\def\[{\begin{equation}}
\def\]{\end{equation}}

\def\/{\over}

\def\os{\oplus}

\def\Re{\mathop{\rm Re}\nolimits}
\def\Im{\mathop{\rm Im}\nolimits}

\begin{document}
\numberwithin{equation}{section}
\numberwithin{theorem}{section}
\numberwithin{proposition}{section}
\numberwithin{definition}{section}

\date{\today}

\title{Quasi-normal modes for de~Sitter-Reissner-Nordstr{\"o}m black holes}

\author{
Alexei Iantchenko
\begin{footnote}
{Department of Materials Science and Applied Mathematics, Faculty of Technology and Society, Malm\"{o} University, SE-205 06 Malm\"{o}, Sweden, email: ai@mah.se }
\end{footnote}
}

\maketitle

\begin{abstract}
The quasi-normal modes for black holes are the  resonances for the scattering of incoming waves by black holes. Here we consider scattering of massless uncharged Dirac fields propagating in the outer region of  de Sitter-Reissner-Nordstr{\"o}m black hole, which is spherically symmetric charged exact solution of the Einstein-Maxwell equations.  Using the spherical symmetry of the equation and restricting to a fixed  harmonic the problem is reduced to a scattering problem for the 1D massless Dirac operator on the line. The resonances for the problem are related to the resonances for  a certain semiclassical Schr{\"o}dinger operator with exponentially decreasing positive potential. We give exact relation between the sets of Dirac and  Schr{\"o}dinger resonances.   The asymptotic distribution of the resonances is close to the lattice of pseudopoles  associated to the non-degenerate maxima of the potentials. 

Using the techniques of quantum Birkhoff normal form we give the complete asymptotic formulas for the resonances. In particular, we calculate the first three leading terms in the expansion.    Moreover, similar results are obtained for  the de Sitter-Schwarzschild quasi-normal modes, thus improving the result of S{\'a} Barreto and Zworski in \cite{SaBarretoZworski1997}.

\end{abstract}

%%%%%%%%%%%%%%%%%%%%%%%%%%%%%%%%%%%%%%%%%%%%
%% MAINMATTER
%%%%%%%%%%%%%%%%%%%%%%%%%%%%%%%%%%%%%%%%%%%%

\section{Introduction}
Quasi-normal modes (QNM) of a black hole are defined as proper solutions of the perturbation equations belonging to certain complex characteristic frequencies
(resonances) which satisfy the boundary conditions appropriate for purely ingoing waves at the event horizon and purely outgoing waves at infinity \cite{ChandrasekharDetweller1975}. It is generally believed that QNMs carry unique footprints to directly identify the black hole existence.
Through the QNMs, one can extract information of the physical parameters of the black hole --- mass, electric charge, and angular momentum --- from the gravitational wave signal by fitting the observed quasi-normal frequencies to those predicted from the mathematical analysis. The subject has become very popular for the last few  decades including the development of   stringent mathematical theory of QNMs (see \cite{DafermosRodnianski2007} and references given there.) For the physics review we refer to \cite{KokkotasSchmidt1999} and more recent \cite{Bertietal2009}.

 Thanks to the work of S{\'a} Barreto and Zworski \cite{SaBarretoZworski1997}, we have a very good knowledge of the localization of QNMs for the wave equation on the
de Sitter-Schwarzschild metric. In Regge-Wheeler coordinates the problem is reduced to the scattering problem for the  Schr{\"o}dinger equation on the line with exponentially decreasing potential.
In the  Schwarzschild  case (zero cosmological constant, which corresponds to asymptotically flat Universe) the Regge-Wheeler potential is only polynomially decreasing and the method does not work due to the possible accumulation of resonances at the origin. A non-zero cosmological constant is needed in order to apply results of \cite{MazzeoMelrose1987} and \cite{Guillarmou2005}, and to define an analytic continuation of the resolvent in a proper space of distributions.

Later, work \cite{SaBarretoZworski1997} was complemented by the paper of Bony and H{\"a}fner \cite{BonyHafner2008}, where the authors considered the local energy decay for the wave equation on the
de Sitter-Schwarzschild metric and proved expansion of the solution  in terms of resonances.

 Here we consider scattering of massless uncharged Dirac fields propagating in the outer region of  de Sitter-Reissner-Nordstr{\"o}m black hole, which is spherically symmetric charged exact solution of the Einstein-Maxwell equations.  We refer to \cite{DaudeNicoleau2011} for detailed study in this background including complete time-dependent scattering theory. We shall use expression obtained in these papers as the starting point of our study. The considered massless Dirac fields are represented by 2-components spinors $\psi$ belonging to the Hilbert space $L^2(\R\times \S^2;\,\C^2)$ which satisfy the evolution equation (Eq.(1.5) in \cite{DaudeNicoleau2011})
\[\lb{1.5}i\partial_t\psi=\left(\sigma_3D_x+\alpha(x)D_{\S^2}\right)\psi,\] where $\sigma_3={\rm diag}(1,-1),$ $D_x=-i\partial_x$ and $D_{\S^2}$ denotes the Dirac operator on the 2D-sphere $\S^2.$ The potential $\alpha$ is defined in  (\ref{a2}) and contains all the information of the metric through the function $F$. Moreover, $\alpha(x)$ decreases exponentially at both infinities, see (\ref{assa}). Note that Dirac operator  \[\label{defD}\cD^{\rm dSRN}=\sigma_3D_x+\alpha(x)D_{\S^2}\]  in the exterior region of  \dSRN black hole takes the same form as a representation of Dirac operator $\cD_\sigma$ on the so called Spherically Symmetric Asymptotically Hyperbolic Manifolds $\Sigma=\R_x\times\S^2_{\theta,\varphi}$ (see \cite{Daudeetal2013}) equipped with the Riemannian metric
$$\s=dx^2+\alpha^{-2}(x)d\omega^2,$$ where $d\omega=d\theta^2+\sin^2\theta d\varphi^2$ is the Euclidean metric on  $\S^2.$  The assumptions on the function  $\alpha(x)$ --- that determines completely the metric ---\\ are $\alpha\in C^2(\R),$ $\alpha >0,$ and
\begin{align} 
\begin{aligned}
&\exists\,\,\alpha_\pm >0,\,\,\pm\kappa_\pm <0\qq\mbox{such that} \label{assa}\\ 
&\alpha(x)=\alpha_\pm e^{\kappa_\pm x}+{\mathcal O}\left( e^{3\kappa_\pm x}\right),\\
&\alpha'(x)=\alpha_\pm \kappa_\pm e^{\kappa_\pm x}+{\mathcal O}\left( e^{3\kappa_\pm x}\right)\qq\mbox{as}\,\,x\rightarrow\pm\infty.
\end{aligned}
\end{align}

 Under these assumptions, $(\Sigma,\sigma)$ is clearly a spherically symmetric Riemannian manifold with two asymptotically hyperbolic ends $\{x=\pm\infty\}$ and the metric $\sigma$ is asymptotically a small perturbation of the ``hyperbolic like'' metrics
$$ \sigma_\pm=dx^2+e^{-2\kappa_\pm}d\omega_\pm^2,\qq x\rightarrow\pm\infty,$$ where $d\omega_\pm^2=1/(\a_\pm^2)d\omega^2$ are fixed metrics on $\S^2.$ Hence, the sectional curvature of $\sigma$ tends to the constant negative values $-\kappa_\pm^2$ on the corresponding ends $\{x\rightarrow\pm\infty\}.$ 

Such spherically symmetric manifolds are very particular cases of the much broader class of asymptotically hyperbolic manifolds (see references in \cite{Daudeetal2013}). We mention also \cite{Vasy2013} for a very general analysis of meromorphic continuation for de Sitter black holes and perturbations.

The analytically extended resolvent of Dirac operator $\cD$ on asymptotically hyperbolic manifolds was described in  \cite{Guillarmouetal2010} using the  parametrix construction extending the ideas from \cite{MazzeoMelrose1987} and  \cite{Guillarmou2005}. 

The massless Dirac operator on $(\Sigma,\sigma)$ $\cD_\sigma=\sigma_3D_x+\alpha(x)D_{\S^2}$  is self-adjoint on the Hilbert space $\cH=L^2(\Sigma;\C^2)$ and has absolutely continuous spectrum. Thus one can define its resolvent   in two ways
$$\cR_+(i\epsilon):=(\cD_\sigma -i\epsilon)^{-1},\qq \cR_-(i\epsilon):=(\cD_\sigma +i\epsilon)^{-1},\qq \epsilon >0,$$ as analytic families of bounded operators on $\cH.$

 From  \cite{Guillarmouetal2010}, Theorem 1.1, it follows that the resolvents $$\cR_\pm(\lambda):\qq C_0^\infty (\Sigma;\C^2)\,\,\mapsto\,\,C^\infty (\Sigma;\C^2)$$ have meromorphic continuation to $\lambda\in\C$ with isolated poles of finite rank.

These properties  can be transmitted  to the operator $ \cD^{\rm dSRN}$ using its identification with a representation of   $\cD_\sigma$ as in \cite{Daudeetal2013}, Eq.(1.4).  Dirac operator $\cD^{\rm dSRN}$ is self-adjoint on $\cH:=L^2(\R\times\S^2, dxd\omega;\C^2),$  its spectrum is purely absolutely continuous and is given by $\R.$ 

The Riemann surface of the resolvent of the Dirac operator $\cR^{\rm dSRN}(\lambda):=(\cD^{\rm dSRN}-\lambda)^{-1}$ consists of two disconnected sheets $\C.$ We will adopt a convention that $\cR^{\rm dSRN}(\lambda)$ is originally defined on $\C_+$ with meromorphic continuation to $\C_-$(which corresponds to the choice of $\cR_+$ above). 
The {\em resonances} or {\em quasi-normal frequencies} are the poles in $\C_-$ of  a meromorphic continuation  of  the cut-off resolvent  
$$\cR_\chi^{\rm dSRN}(\lambda)=\chi(\cD^{\rm dSRN}-\lambda)^{-1}\chi,\qq \chi\in C_0^\infty(\R;\C^2),$$ from the upper half-plane to $\C.$

Note that equivalently we can consider the resolvent on the  lower half plane $\C_-$ and obtain a meromorphic continuation to $\C_+$ (which corresponds to the choice of $\cR_-$ above).

We consider  the scattering of massless uncharged Dirac waves towards the two ends 
 $\{x\rightarrow\pm\infty\}$ in the context of \dSRN black holes.

We show that the situation is similar to the scattering problem for the wave equation on the \dSS metric. The scattering phenomena there (see  \cite{SaBarretoZworski1997}, Eq.(4.2)) are governed by the Schr\"odinger operator \[\lb{defP}\cP^{\rm dSS}=D_x^2+\alpha^2[\Delta_\omega+2\alpha \alpha' r^3+2\alpha^2 r^2]\] as operator in $(x,\omega)$ on $L^2(\R\times \S^2;\,\C),$ where $\alpha$ is as in (\ref{a2}) but with $Q=0$ and $r=r(x)$ via Regge-Wheeler transformation (\ref{Regge}). Here $\Delta_\omega$ is the (positive) Laplacian on $\S^2.$ The resonances for \dSS black holes are defined as the poles $\lambda\in \C_-$ of the meromorphic continuation of the cut-off resolvent
$$\cR_\chi^{\rm dSS}(\lambda)=\chi(\cP^{\rm dSS}-\lambda^2)^{-1}\chi,\qq \chi\in C_0^\infty(\R),$$ from $\C_+$ to $\C.$

 The resonances are approximated by the lattice associated to the trapped set which is a sphere of partially hyperbolic orbits ---  {\em photon sphere\,\,}(see \cite{Gerard1988}, \cite{GerardSjostrand1987}). Due to radial symmetry, after separation of variables and a Regge-Wheeler transformation the problem is reduced to a family of one-dimensional Schr\"odinger operators on a line with potentials exponentially decaying at infinity and having unique non-degenerate maxima. Using the inverse of the angular momentum as a semiclassical parameter, the result of  \cite{Sjostrand1986} gives the leading order in the expansion of resonances (see  \cite{SaBarretoZworski1997}). 

We show that resonances for \dSRN black holes can be obtained as solutions of one-dimensional Schr\"odinger equations with similar properties as in \dSS case. Moreover, using the method of quantum Birkhoff normal form (as in \cite{Iantchenko2007}, \cite{Iantchenko2008}) we obtain complete asymptotic 
expansions in both \dSS and \dSRN cases.

From the physicists point of view, the quasi-normal modes for Reissner-Nordstr{\"o}m black holes were calculated numerically in \cite{BertiKokkotas2003}, \cite{WuZhao2004} (massless case), \cite{ChangShen2007} (massive case) and \cite{Jing2004} (de Sitter variant of the massless case). 
Note that the authors treated the Dirac resonances exactly as solutions of the Schr{\"o}dinger equation similar to (\ref{Schr}) (see also \cite{Chandrasekhar1983}, \cite{Chandrasekhar1980}). Our main result, Theorem \ref{th-dSRN}, shows a different  point of view and gives exact relation between Schr{\"o}dinger and Dirac resonances.  Indeed, due to the symmetry of the equation, the set of  non-zero Schr{\"o}dinger resonances consists of two sets interposed:
the set of Dirac resonances and its mirror image  with respect to the imaginary axis.

Our reason to study massless and uncharged fields is that the resulting Dirac operator 
coincides with a representation of a $\cD$ on the Spherically Symmetric Asymptotically Hyperbolic Manifolds $\Sigma$ as above and the global properties of its resolvent are already known thanks to  \cite{Guillarmouetal2010}. Moreover, the one-dimensional massless Dirac operator is a 2-by-2 matrix operator with exponentially decreasing potential, whereas in the massive charged case it must be a 4-by-4 matrix operator with the potential decreasing exponentially to some non-zero constants at infinities (see \cite{DaudeNicoleau2010} and \cite{Gobin2015}). For the massless uncharged fields the Dirac operator 
has supersymmetric structure (see  \cite{Chandrasekhar1980}, \cite{Thaller1992} and \cite{Jing2004}) and
 has a nice relation to a Schr{\"o}dinger operator similar 
to that appearing  in scattering problem for the wave equation in \dSS metric (see  \cite{SaBarretoZworski1997}). As the last problem is well studied, we can easily transmit  many already existing results to the Dirac case, and  apply the Birkhoff normal form construction. Note also that the formulas obtained in this paper for the massless uncharged case indicate what one should expect to get in the general case as it is believed that, due to intense gravitation near the event and cosmological horizons of the black hole, even if the Dirac fields are massive, they propagate asymptotically as in the massless case (see \cite{Gobin2015}, \cite{Iantchenko2016a}).  

Recently,  several new mathematical works on quasi-normal modes in other backgrounds have appeared. We mention few of them.

In \cite{Dyatlov2011} and \cite{Dyatlov2012} Dyatlov studied the slowly rotating Kerr-de Sitter black holes. Due to cylindrical instead of spherical symmetry the problem can no longer be simply reduced to a scattering problem on the line. The quasi-normal modes split in a way similar to the Zeeman effect. Dyatlov also extended \cite{BonyHafner2008} to the rotating black holes and showed the exponential decay of local energy of linear waves orthogonal to the zero quasi-normal mode. Note also the paper \cite{DaudeNicoleau2015}, where the authors extended their inverse scattering results from \cite{DaudeNicoleau2011} to the scattering for massless Dirac fields by the (rotating) Kerr-Newman-de Sitter black holes.
It appears that the techniques of Dyatlov used in \cite{Dyatlov2011}, \cite{Dyatlov2012} can be adapted to the framework of Dirac operators for the Kerr-Newman-de Sitter black holes (see 
\cite{Iantchenko2016a}).

In \cite{Gannot2014} and \cite{Warnick2015} the quasi-normal modes in rather different geometry of Anti-de-Sitter (AdS) black holes are discussed. Such black holes arise in superstring theory via AdS conformal field theory correspondence, that string theory in AdS space is equivalent to conformal field theory in one less dimension (see \cite{Bertietal2009},  \cite{Warnick2015}). The quasi-normal frequencies correspond to the thermalization time scale, which is very hard to compute directly.   Gannot in \cite{Gannot2014} uses a black-box approach to define the quasi-normal modes
after separation of variables and furthermore finds a sequence of quasi-normal frequencies
approaching the real axis exponentially rapidly.   Warnick in \cite{Warnick2015} uses a different approach which applies to asymptotically Anti-de-Sitter black holes and does  not require any separability of the equations
under consideration, nor any real analyticity of the metric. Moreover, the method  can be extended to asymptotically de-Sitter black holes, where it is closely related to approach by Vasy in \cite{Vasy2013}, and  permits consideration
of perturbations which do not vanish on the horizons.

 The present paper is the first one in our project on quasi-normal modes for Dirac fields in black hole geometries. In \cite{Iantchenko2015b} we get an expansion of the solution of the massless Dirac equation in \dSRN metric in terms of resonances and show exponential decay of local energy for compactly supported data, similar to \cite{BonyHafner2008} and \cite{Dyatlov2011}.
The method is based on the relation between the Dirac and the Schr{\"o}dinger operators established in the present paper and the cut-off resolvent estimates from \cite{BonyHafner2008}.

In \cite{Iantchenko2016a} we provide the full asymptotic description  of the quasi-normal modes (resonances) in any strip of fixed width for  Dirac fields in slowly rotating Kerr-Newman-de Sitter  black holes.  The resonances split in a way similar to the Zeeman effect. The method is based on the extension to Dirac operators of techniques applied by Dyatlov  in \cite{Dyatlov2011}, \cite{Dyatlov2012} to the scalar fields  in (uncharged) Kerr-de Sitter black holes.  We show that the mass of the Dirac field does not have effect on the two leading terms in the expansions of resonances. However, contrary to the present paper, in  \cite{Iantchenko2016a}  we were unable to calculate explicitly the leading third term in the expansion.  Indeed, due  to cylindrical instead of spherical symmetry, the angular and the radial parts of the governing equation do not decouple. The problem can no longer be simply reduced to scattering problem for the supersymmetric Dirac operator on the line as  in the non-rotating case considered in the present paper, and explicit calculation of the quantum Birkhoff normal form coefficients is much more challenging.

The author thanks the referees for numerous comments and suggestions.

\section{Definitions and main results}
 In this section we recall the orthogonal decomposition of the  Dirac operator, summarize the properties of  the one-dimensional Dirac operator  and formulate the main results. 

By decomposition (see Section 2.1 in \cite{DaudeNicoleau2011}) of the Hilbert space $\cH=L^2(\R\times\S^2, dxd\omega;\C^2)$ in spin-weighted spherical harmonics $F_m^l,$ $(l,m)\in\cI,$
$$\cI=\left\{(l,m);\,\,l-\frac12\in\N,\,\,l-|m|\in\N\right\},\qq\cH=\bigoplus_{(l,m)\in\cI}\cH_{l,m},$$
where $\cH_{l,m}$ is identified with $L^2(\R;\C^2),$
 we obtain the orthogonal decomposition for the Dirac Hamiltonian $\cD^{\rm dSRN}$
$$\cD^{\rm dSRN}=\bigoplus_{(l,m)\in\cI}\cD^{l,m},\qq \cD^{l,m}:=\cD^{\rm dSRN}_{|\cH_{l,m}}=\s_3D_x-\left(l+\frac12\right) \a(x)\s_1,$$ where
the one-dimensional Dirac operator $\cD^{l,m}$ does not depend on index~$m.$ 

Now, the scattering of massless charged Dirac fields in de Sitter-Reissner-Nordstr{\"o}m black holes is described (see \cite{DaudeNicoleau2011}, Eq.(2.14)) by the scattering on the line for the massless Dirac system
\begin{equation}\label{DiracSystem}
\begin{aligned}
&\left[ \s_3D_x-n \a(x)\s_1\right]\psi=\lambda\psi,\\
&n=l+\frac12\in\N,\qq \s_3=\left(
         \begin{array}{cc}
           1 & 0 \\
           0 & -1 \\
         \end{array}
       \right)\qq \s_1=\left(
         \begin{array}{cc}
           0 & 1 \\
           1 & 0 \\
         \end{array}
       \right),
\end{aligned}
\end{equation}
which is a special form of Zakharov-Shabat system (see \cite{IantchenkoKorotyaev2013} with $q=\linebreak -n \a(x)\in\R$).  The potential $\a(x)$ is given by
\begin{equation}\label{a2}
\a^2(x)=\frac{F(r(x))}{r^2(x)},\qq F(r)=1-\frac{2M}{r}+\frac{Q^2}{r^2}-\frac{\Lambda}{3}r^2,
\end{equation}
where $M>0,$ $Q\in\R$ are the mass and the electric charge of the black hole respectively, $\Lambda >0$ is the cosmological constant.
The equation (\ref{DiracSystem}) is expressed by means of Regge-Wheeler coordinate $x$ related to the original radial coordinate $r$ by means of the equation
\[\lb{Regge}\frac{dx}{dr}=\frac{1}{F(r)}.\]
 We suppose that $Q^2 <\frac98M^2$ and $\Lambda M^2$ is small enough. Then the function $F(r)$ has four real zeros
 $$r_n < 0< r_c<r_-<r_+.$$ The sphere $\{r=r_c\}$ is called the Cauchy horizon, whereas the spheres $\{r=r_-\}$ and $\{r=r_+\}$ are the event and cosmological horizons respectively.

The Regge-Wheeler radial variable $x$ is given explicitly for $r_-<r<r_+$ by
\begin{align}\lb{1.3DN}
x&=\frac{1}{2\kappa_n}\ln (r-r_n)+\frac{1}{2\kappa_c}\ln (r-r_c)\\
\notag &\quad +\frac{1}{2\kappa_-}\ln (r-r_-)+\frac{1}{2\kappa_+}\ln (r_+-r)+c,
\end{align}
where $c$ is any constant of integration and the quantities $\kappa_j,$ $j=n,c,-,+$ are defined by
 $$\kappa_n=\frac12 F'(r_n),\quad  \kappa_c=\frac12 F'(r_c),\quad \kappa_-=\frac12 F'(r_-),
 \quad\kappa_+=\frac12 F'(r_+).$$

  We consider scattering in the exterior region $\{ r_- <r<r_+\},$ where we have \[\lb{expdec}  \a(x)\sim \a_\pm e^{\kappa_\pm x}\qq\mbox{as}\,\,x\rightarrow\pm\infty\,\,\mbox{or}\,\,r\rightarrow r_\pm,\] where
 $\kappa_- >0,$  $\kappa_+<0$ are surface gravities at  event and cosmological horizons respectively,  $\alpha_\pm$ are fixed constants depending on the parameters of the black hole.
% Note that $V_0(x)=\a^2(x)$ has a non-degenerate maximum at $x_0=x(r_0),$
 % ${\displaystyle r_0=3M/2+\sqrt{\left(3M/2\right)^2-2Q^2}.}$
% and generate a sphere of hyperbolic orbits for Hamiltonian in (\ref{DiracSystem}).

It is well known (see \cite{DEGM})   that the operator $\s_3D_x-n \a(x)\s_1$ acting in $L^2(\R )\os L^2(\R )$ is
self-adjoint and its spectrum  is purely absolutely
continuous and is given by the set $\R$. In \cite{IantchenkoKorotyaev2013} we studied resonances of such operators in the case of compactly supported potential $-n\a(x).$ Then the outgoing  solutions (Jost solutions) have analytic continuation from the upper half-plane $\C_+$ to the whole complex plane $\C$ and resonances are the zeros in $\C_-$ of the Wronskian for the Jost solutions or, equivalently, the poles in    $\C_-$ of the analytic continuation of the cut-off resolvent. For non-compactly supported exponentially decreasing potential $q(x)=-n \a(x)$ satisfying (\ref{expdec}) such method of analytic continuation  is possible in a strip $\{\lambda\in\C;\,\,\Im\,\lambda>-\epsilon\}$ for some $\epsilon >0$ (see \cite{Froese1997}). In order to calculate resonances in a larger domain (a sector) one uses  the method of complex scaling under the condition that the potential $\alpha$ admits holomorphic extension in a conic \neigh{} of a real axis and exponentially decays there. This property is shown  below in Section \ref{s-Bar}, Proposition  \ref{Prop4.1}.

It is well-known that different definitions give rise to the same set of resonances in the domains where both definitions are applicable (see \cite{HelfferMartinez1987}).

We use that  (\ref{DiracSystem}) can be written in the semiclassical way as
$$
\cD_{-\alpha}\psi\equiv\left[ h\s_3D_x-\alpha(x)\s_1\right]\psi=z \psi,\,\, z=\lambda/n,$$ 
with ``Planck constant'' $ h=1/n.$ 
We denote the set of resonances for $\cD_{-\alpha}=h\s_3D_x-\alpha(x)\s_1$ by ${\rm Res}\,(\cD_{-\alpha})\subset \C_-.$
Note the following symmetry property of the Dirac operator $\cD_{-\alpha}$ with real-valued $\alpha:$  $$\lambda\in{\rm Res}\,(\cD_{-\alpha})\,\,\Leftrightarrow\,\,-\overline{\lambda}\in{\rm Res}\,(\cD_{\alpha}).$$

We consider also the Schr\"odinger operator  \[\lb{Schr}P=h^2(D_x)^2+V_h(x),\qq V_h(x)=\a^2(x)+h\a'(x).\] We say that $\lambda\in\C_-$ is a resonance for $P$ if, for some function $\chi\in C_0^\infty(\R),$ $\lambda$ is a  pole  of meromorphic continuation of the cut-off resolvent
$\chi(P-\lambda^2)^{-1}\chi.$
We denote the set of resonances of $P$ by ${\rm Res}\,(P).$ The set of resonances is invariant under the change of sign $\alpha$ $\mapsto$ $-\alpha$ and invariant under  the reflection ${\rm S}$ with respect to $i\R:$
$\lambda\in{\rm Res}\,(P)\,\,\Leftrightarrow\,\,-\overline{\lambda}\in{\rm Res}\,(P).$

For a  set of points $\sigma=\{\lambda_j\}\in\C_-$ we denote the mirror image with respect to $i\R$ by \[\lb{mirror}\sigma^{\rm S}:=\{-\overline{\lambda_j}\}\in\C_-.\]

In Section \ref{s-Red} we show  the following relation between resonances for $\cD_{\pm \alpha}$ and $P$ (see Lemma \ref{L-UnD}):
$${\rm Res}\,(P)\setminus\{0\}={\rm Res}\,(\cD_{-\alpha})\cup{\rm Res}\,(\cD_{\alpha})={\rm Res}\,(\cD_{-\alpha})\cup{\rm Res}^{\rm S}\,(\cD_{-\alpha}).$$

The principal symbol of the potential in (\ref{Schr}) $V_0(x)=\a^2(x)$ has a non-degenerate maximum at $x_0=x(r_0),$ where
$$
 r_0=\frac{3M+\sqrt{(3M)^2-8Q^2}}{2},\qq V_0(x_0)=r_0^{-4}\left(Mr_0-Q^2-\frac{\Lambda}{3}r_0^4,\right).$$ 
The derivatives of the potential at $x=x_0$ are given by 
\begin{align}V_0''(x_0)=&\left(\frac{4Q^2}{r_0^2}-2\right)V_0^2(r_0)=-2\left(\frac{3M}{r_0}-\frac{4Q^2}{r_0^2}\right)V_0^2(x_0)\lb{somef}\\
V_0'''(x_0)=&\frac{4}{r_0}\left(11Mr_0-18Q^2-8Mr_0^3+12Q^2r_0^2+\frac43\Lambda\left[r_0^4-r_0^6\right]\right)V_0^3(x_0).\nonumber
\end{align}
In this special one-dimensional case the result by Sj{\"o}strand  \cite{Sjostrand1986} implies that the resonances associated to the non-degenerate maximum of the principal symbol $V_0(x)$ of the potential, barrier top resonances, are close to the string of pseudopoles with constant real part.

%\[\lb{somef}V_0''(x_0)=\left(\frac{dx}{dr}\right)^{-2}\left.\frac{d^2V_0}{dr^2}\right|_{r=r_0}=-2\left(\frac{3M}{r_0}-\frac{4Q^2}{r_0^2}\right)V_0^2(x_0).
%\]

%Here, $$\left(\frac{dx}{dr}\right)^{-2}=F^2(r),\qq V_0=\frac{F(r_0)}{r_0^2},\qq \left.\frac{d^2V_0}{dr^2}\right|_{r=r_0}=\frac{2}{r_0^6}\left(-3Mr_0+4Q^2\right). $$

% We have also \begin{align*}
%V_0''(x_0)=&\left(\frac{4Q^2}{r_0^2}-2\right)V_0^2(r_0),\qq \frac{3M}{r_0}-\frac{4Q^2}{r_0^2}= 1-\frac{2Q^2}{r_0^2},\\
%V_0'''(x_0)=&\frac{4}{r_0}\left(11Mr_0-18Q^2-8Mr_0^3+12Q^2r_0^2+\frac43\Lambda\left[r_0^4-r_0^6\right]\right)V_0^3(x_0).\end{align*}

Note that 
resonances (quasi-normal modes) for an operator similar to (\ref{Schr})    were  mathematically studied in \cite{SaBarretoZworski1997} and \cite {BonyHafner2008} in  the context of \dSS black holes. The authors of \cite{SaBarretoZworski1997} give two leading terms in the asymptotic expansions of resonances. We show that similar results also hold for the \dSRN resonances. Namely, we show that in semiclassical limit $h=1/(l+1/2)\rightarrow 0$ the resonances are close to the lattice of pseudopoles. Moreover, using  the method of  semiclassical (or quantum)  Birkhoff normal form (abbreviated qBnf, see \cite{KaidiKerdelhue2000} and \cite{Iantchenko2008}) we get the complete asymptotic expansions for the resonances both in \dSRN and \dSS cases. 

Now, using  the explicit reconstruction procedure of the qBnf as in  \cite{CdVerdiereGuillemin2011} we get explicit formulas for the next (third) order terms in the expansions of resonances.

The main result of this paper is the following theorem.

\begin{theorem}[\dSRN resonances]\lb{th-dSRN} 
Let $$\Omega_{C}=\{\lambda\in\C_-;\,\,\Im\lambda>-C,\,\,\Re\lambda >K,\,\,\Im\lambda >-\theta |\Re\lambda| \}.$$ 
%For any $N\in\N$ and for all $\theta>0,$ $\delta >0$ and $C>0$ there exist $K>0$ and %$m=m(N)\in\N$
Fix a number  $N\in\N.$ Then there exist   $K>0,$  $\theta>0,$  $r\in\N$ and functions $f_j=f_j(2k+1)={\mathcal O}\left( (2k+1)^j\right),$ $k\geq 0,$ $j=1,\ldots,r,$ polynomial in $2k+1$ of order $\leq j,$ such  that
 for any $C>0$  there exists an injective map, $b_N$ from the set of pseudo-poles
\begin{align*}\mu_{k,l}^r=&  (l+1/2)\left(z_0+\frac{f_1(2k+1)}{l+1/2}+\frac{f_2(2k+1)}{(l+1/2)^2}+\cdots+\frac{f_{r+1}(2k+1)}{(l+1/2)^{r+1}}\right),\\& \,\,l+1/2\in\N,\,\,k\in\N_0,
\end{align*}
into the set of resonances \[\lb{UnD}{\rm Res}\,(\cD^{\rm dSRN})\cup{\rm Res}^{\rm S}\,(\cD^{\rm dSRN}),\qq \cD^{\rm dSRN}=\sigma_3D_x-\alpha(x)D_{\S^2},\]  such that all the resonances in $\Omega_{C}$ are in the image of $b_N$ and for $b_N(\lambda)\in\Omega_{C},$
$$b_N(\lambda)-\lambda={\mathcal O}\left(|\lambda |^{-N}\right). $$ Here  $\{\cdot\}^{\rm S}$ denotes the mirror reflection of the set $\{\cdot\}\!\in\!\C_-$ in $i\R$ (see (\ref{mirror})) and 
\begin{align*}&z_0=\alpha(x_0),\quad \omega=\left(\frac12|V_0''(x_0)|\right)^\frac12,\\
&f_1=- \frac{i\omega}{2 z_0}  (2k+1),\\
&f_2=-\frac{i\omega}{2 z_0}  (2k+1)\left[-\frac{1}{4iz_0^2}\omega(2k+1)+\frac{1}{2i} b_{0,2}(2k+1)+b_{1,2} \right],\\ 
&b_{0,2}=\frac{15}{4\cdot 12^2}\frac{(V_0'''(x_0))^2}{\omega^5}+\frac{V_0''''(x_0)}{32\omega^3},\qq b_{1,2}=\frac{1}{8z_0^3}-\frac{3}{8z_0\omega^2}V_0'''(x_0).
 \end{align*}

 The resonance in ${\rm Res}\,(\cD^{\rm dSRN})$ corresponding to pseudopole $\mu_{k,l}^r$ has multiplicity $2l-1.$

\end{theorem}

\noindent{\bf Remark 1.}
From Theorem \ref{th-dSRN}  we get in the leading order that resonances in $\Omega_{C}$ are approximated by pseudopoles
\begin{align}
\mu_{k,l}&=z_0(l+1/2)-i\left(\frac{\omega}{ z_0}\right)  \left(k+\frac{1}{2}\right)\lb{leading}\\
\notag &\quad -\left(\frac{\omega}{ z_0}\right) \frac{(k+1/2)}{(l+1/2)}\left[-\frac{1}{4z_0^2}\omega(2k+1)+\frac{1}{2} b_{0,2}(2k+1)+ib_{1,2} \right]\\
\notag &\quad +{\mathcal O}((l+1/2)^{-2}), 
\end{align}
where 
\begin{align*}
&z_0=\left(\frac{M}{r_0^3}-\frac{Q^2}{r_0^4}-\frac{\Lambda}{3}\right)^\frac12,\qq \frac{\omega}{ z_0}=\left(\frac{3M}{r_0}-\frac{4Q^2}{r_0^2}\right)^\frac12\left(\frac{M}{r_0^3}-\frac{Q^2}{r_0^4}-\frac{\Lambda}{3}\right)^\frac12,\\
&\omega=\left(\frac{3M}{r_0}-\frac{4Q^2}{r_0^2}\right)^\frac12 z_0^2(x_0).
\end{align*}
The slowest damped mode as $l\rightarrow \infty$ (the leading terms (\ref{leading}) for $k=0$)
\begin{align*}
&\Re \mu_{0,l}\approx \left(l+\frac12\right)\Omega_0,
\qq \Im  \mu_{0,l}\approx -\frac12 \Omega_0\left[\frac{3M}{r_0}-\frac{4Q^2}{r_0^2}\right]^\frac12,\\
&\Omega_0=\left[\frac{M}{r_0^3}-\frac{Q^2}{r_0^4}-\frac{\Lambda}{3}\right]^\frac12,
\end{align*}
where
$ \Omega_0$ is frequency of the unstable circular null geodesics with radius $r_0,$ 
was obtained before for $\Lambda=0$ (see formulas (4.4.5) and (4.4.6) in \cite{FrolovNovikov1998}, formula (93) in \cite{Bertietal2009}) using WKB methods for $l>>1.$ So our result mathematically justifies this approach and provides lower order corrections.

\medskip
\noindent {\bf Remark 2} (\dSS resonances). Similar result is also  true for 
 the \dSS resonances. Namely,
Theorem \ref{th-dSRN} is valid for ${\rm Res}\,(\cP^{\rm dSS})$   in place of (\ref{UnD}), where $\cP^{\rm dSS}$ is given in (\ref{defP}), 
after putting $Q=0$ in the definition of the function $\alpha$  in (\ref{a2}). Then  
$$z_0^2=\alpha^2(x_0)=\frac{1-9\Lambda M^2}{3^3M^2}=\left(\frac12|V_0''(x_0)|\right)^\frac12=\omega,$$ $f_1$ is given by the same formula as above, $b_{0,2}$ is the same,
$$f_2=-\frac{\omega}{ z_0}  \left[\frac18+\frac{1}{4} b_{0,2}(2k+1)^2- b_{2,0} \right],\qq b_{2,0}=\frac{(V_0'''(x_0))^2}{12^2\omega^5}+9\omega M^6.$$ Then we get in the leading order that resonances in $\Omega_{C}$ are approximated by pseudopoles
\begin{align*}
\mu_{k,l}&=\omega^\frac12\left( (l+l/2) -i(k+1/2)\right)\\
&\quad -\omega^\frac12(l+1/2)^{-1}\left(\frac18+\frac{1}{4} b_{0,2}(2k+1)^2- b_{2,0}\right)\\
&\quad +{\mathcal O}\left(l+1/2\right)^{-2}.\end{align*}
Here, the first two terms are well-known (see \cite{SaBarretoZworski1997}) and coincide with
(\ref{leading}) after putting $r_0=3M,$ $Q=0.$

The slowest damped mode as $l\rightarrow \infty$ (the leading terms  for $k=0$) is
\begin{align*}
&\Re \mu_{0,l}\approx \left(l+\frac12\right)\Omega_0,
\qq \Im  \mu_{0,l}\approx -\frac12 \Omega_0,\\
&\Omega_0=\left[\frac{1-9\Lambda M^2}{3^3M^2}\right]^\frac12,\qq r_0=3M.
\end{align*}

The resonance corresponding to pseudopole $\mu_{k,l}$ has multiplicity $2l+1.$

\medskip
\noindent {\bf Remark 3.} The value of cosmological constant $\Lambda$ does  not have a physical effect on the quasi-normal modes since gravitational waves are generated in a \neigh{} of black hole. 
But asymptotically hyperbolic geometry  for $\Lambda >0$ makes the mathematical definition of quasi-normal modes much easier by eliminating the polynomial fall-off for waves which occurs for asymptotically flat black holes ($\Lambda=0$).
However, the formal expansions in  this paper remain valid even in the case of zero cosmological constant.
\vspace{5mm}

The paper is organized as follows. In Section \ref{s-Red}, we show how resonances for the Dirac operator can be calculated from the resonances for a certain Schr{\"o}dinger operator. In Section \ref{s-Bar}, we consider the asymptotic expansions for the resonances generated by the non-degenerate maximum of the potential --- barrier top resonances. In Section  \ref{ss-Bar}, we consider the analytic properties of the potential and by complex scaling show relation between resonances and pseudopoles. In Section \ref{ss-Exp}, we apply the method of \cite{CdVerdiereGuillemin2011} to our  Schr{\"o}dinger operator in order to get explicit formulas for the coefficients of the qBnf.
 In Sections \ref{s-dSRN} and \ref{s-dSS}, we prove Theorem \ref{th-dSRN}  and  Remark 2.

\section{Reduction to Schr{\"o}dinger equation} \lb{s-Red}
We consider Dirac operator $\cD=\cD_q=\cD_0+V:=-i\s_3\partial_x+q\s_1$ and  the\linebreak  Dirac equation for a vector-function $f(x)$
\[\lb{ZSs}\begin{aligned}
&-i\s_3f'+q\s_1  f=\lambda f, \ \ \ \lambda \in \C,\\
&f(x)= \ma f_1\\ f_2\am=f_1(x)e_++f_2(x)e_-,  \ \ \ e_+= \ma 1\\0\am,
   \   e_-= \ma 0\\1\am,\end{aligned}
\]
where $f_1, f_2$ are  the functions of $x\in\R $. Note that (\ref{ZSs}) is also known as (a special case of)
Zakharov-Shabat system. Inverse scattering theory for Zakharov-Shabat systems plays an important role for the investigation of NLS (see \cite{DEGM}).
Here $q(x)=-n\a(x)<0$ for the \dSRN black hole satisfying (\ref{expdec}): $\a(x)\sim \a_\pm e^{\kappa_\pm x}$ as $x\rightarrow \pm\infty,$ $\pm\kappa_\pm <0.$ 
The Jost solutions $\p_{\pm}, \vp_{\pm}, $ of (\ref{ZSs})  defined via
the following asymptotics
$$
 \vp^{\pm}(x,\lambda )\sim e^{\pm i\lambda x}e_{\pm},\ \ \ \  x\rightarrow -\infty;\qq \p^{\pm}(x,\lambda )\sim e^{\pm i\lambda x}e_{\pm},\ \ \ \  x\rightarrow +\infty  ,
$$
satisfy the identity
$$\vp^-(x,\lambda )=b(\lambda )\p^+(x,\lambda )+a(\lambda )\p^-(x,\lambda ).$$
Functions $a(\lambda),$ $b(\lambda)$ are analytic in $\C_+$ (see \cite{IantchenkoKorotyaev2013}) and for non-compactly supported potential exponentially decreasing at both infinities they have analytic continuation  over continuous spectrum in a strip  $\{\lambda\in\C;\,\,\Im\,\lambda>-\epsilon\}$ for some $\epsilon >0.$ In this strip the resonances are the  zeros of $a(\lambda).$  Alternatively, for real analytic $q,$
 the resonances can be obtained by the method of complex scaling (\cite{Weder1973}, \cite{Seba1988a} and \cite{Cueninetal2013})  by defining analytic continuation of the resolvent $(\cD-\lambda)^{-1}$ from $\C_+$ into $\C.$  The resonances $\lambda\in\C_-$ are the poles of meromorphic continuation of 
 $$\chi(\cD-\lambda)^{-1}\chi,\qq \chi\in C_0^\infty(\R).$$ Recall that the Riemann surface of the resolvent for the massless Dirac operator consists of two disconnected sheets $\C$ (see \cite{IantchenkoKorotyaev2013})  and we agreed to consider all functions and the resolvent in $\C_+$ and to obtain analytic continuation to $\C.$ The lower half-plane $\C_-$ is the ``unphysical sheet'' for the Dirac operator.
  Now, for the Dirac operator
$$\cD=-i\s_3f'+q\s_1=  \ma -i\partial_x & q\\q & i\partial_x\am$$  we 
consider also its square $$\cD^2=-I_2\partial^2_x +\ma q^2&-iq'\\iq'&q^2\am,$$ which is matrix Schr{\"o}dinger operator. Operator $\cD^2$ is self-adjoint in  $L^2(\R )\os L^2(\R )$ and unitary equivalent to
$$U\cD^2U^{-1}=\ma \cP_-&0\\
0&\cP_+\am,\qq \cP_\pm=-\partial_x^2 +q^2\pm q'.$$
Here, $$U=\frac{i}{\sqrt{2}} \ma 1 & i\\1 & -i\am,\qq U^{-1}=-\frac{i}{\sqrt{2}} \ma 1 & 1\\-i & i\am .$$
The resolvents $(\cD^2-\lambda^2)^{-1},$ $(\cP_\pm-\lambda^2)^{-1}$ are analytic functions on $\C_+$ and admit analytic continuation into $\C.$ The Riemann surface of the Schr{\"o}dinger resolvents (\wrt{} ``frequency'' $\lambda$) consists of ``physical sheet''  $\C_+$ and ``unphysical sheet'' $\C_-.$

The resonances for $\cP_\pm$ are the poles $\lambda\in\C_-$ of meromorphic continuation
of $$\chi(\cP_\pm-\lambda^2)^{-1}\chi,\qq \chi\in C_0^\infty(\R).$$ Note that $\cP_+$ and $\cP_-$ have identical resonances.
The two potentials $q^2\pm q'$  are supersymmetric partners
derived from the same superpotential $q$ (see  \cite{Chandrasekhar1980}, \cite{Thaller1992} and \cite{Jing2004}).

Now, consider the identity 
\begin{align}\lb{SchrDirac}
&\chi(\cD^2-\lambda^2)^{-1}\chi=(2\lambda)^{-1}\left[\chi(\cD-\lambda)^{-1}\chi-\chi(\cD+\lambda)^{-1}\chi\right],\\
\notag &\chi\in C_0^\infty(\R),
\end{align} 
which due to (\ref{expdec}) is well defined in a small strip $\{\lambda\in\C,\,\,0<\Im\lambda<\epsilon\}\in\C_+$ and has meromorphic continuation to $\C,$ whose poles $\lambda\in \C_-$ are the resonances for $\cD^2.$ Here, 
\begin{align*}
&U(\cD^2-\lambda^2)^{-1}U^{-1}=\left(
           \begin{array}{cc}
            (\cP_- -\lambda^2)^{-1}& 0 \\
             0 &  (\cP_+-\lambda^2)^{-1}\\
           \end{array}
         \right),\\
         &\cP_\pm=-\partial_x^2 +q^2\pm q',
         \end{align*}
         and the sets of resonances for $\cD^2$ and $\cP\equiv \cP_+$ coincide.  We denote the set of resonances for the Schr\"odinger operator $\cP$ by ${\rm Res}\,(\cP).$ 

Note the following symmetry property of the resonances for $\cP:$
 \[\lb{symSch}\lambda\in{\rm Res}\,(\cP)\qq\Leftrightarrow\qq -\overline{\lambda}\in{\rm Res}\,(\cP).\]
The set ${\rm Res}\,(\cP)$  is invariant under change of sign $q$ $\mapsto$ $-q.$

Now, we consider two Dirac operators  $\cD_{\pm q}=-i\s_3f'\pm q\s_1$ with the respective resonance sets ${\rm Res}\,(\cD_{\pm q}).$ Note the following symmetry property:
 \[ \lb{symD}\lambda\in{\rm Res}\,(\cD_{q})\qq\Leftrightarrow\qq -\overline{\lambda}\in{\rm Res}\,(\cD_{-q}).\]
Let $\{\cdot\}^{\rm S}$ denote the mirror reflection of the set $\{\cdot\}$ in $i\R$ (see (\ref{mirror})). Identities (\ref{SchrDirac}), (\ref{symSch}) and (\ref{symD}) imply 
\begin{lemma}\lb{L-UnD} The set of non-zero resonances of the Schr{\"o}dinger operator
$\cP=-\partial_x^2+q^2+q' $ has the following decomposition:
$${\displaystyle {\rm Res}\,(\cP)\setminus\{0\}={\rm Res}\,(\cD_{q})\cup{\rm Res}\,(\cD_{-q})={\rm Res}\,(\cD_{q})\cup{\rm Res}^{\rm S}\,(\cD_{q})\in\C_-,}$$ where
$\cD_{\pm q}=-i\s_3f'\pm q\s_1$ are Dirac operators and ${\rm S}$ denotes mirror reflection of a set in $i\R.$
Here we identify the ``unphysical sheet'' $\C_-$ for the  Schr\"odinger operator with  the ``unphysical sheet'' $\C_-$ for the Dirac operator. 
\end{lemma}

\section{Barrier top resonances}\lb{s-Bar}
\subsection{Resonances and pseudopoles}\lb{ss-Bar}
We start by recalling the analytic properties of the radial coordinate $r$ as a function of the Regge-Wheeler variable $x$ (see (\ref{Regge})) and a holomorphic extension of the potential $\alpha$ in (\ref{a2}). For $Q=\Lambda=0$ it was proved in \cite{SaBarretoZworski1997},
Proposition~4.1 (see also \cite{BachelotMotetBachelot1993}, Propositions IV.2 and IV.3 and \cite{Dyatlov2011}, Proposition~4.1).
\begin{proposition}\lb{Prop4.1} Let $F$ be as in  (\ref{a2}). Suppose $Q^2 <\frac98M^2$ and $\Lambda M^2$ is small enough so that   $F(r)$ has four real zeros
 $r_n < 0< r_c<r_-<r_+.$ 
Let the function $x=x(r)$ be defined by $$ x=\int_{r_0}^r\frac{ds}{F(s)},$$ where $r_0\in (r_-,r_+)$ is a fixed number. 

Then the functions $r(x)$ and $\alpha(x)=\sqrt{F(r(x))}/r(x)$ extend to a holomorphic function in a conic neighborhood of the real axis given by $|\arg z|<\theta$ and $\alpha$ satisfies there
\[\lb{bound_conic}|\alpha(z)|\leq C\exp (-|z|/C),\qq \Re z\rightarrow\pm\infty. \]
\end{proposition}
\begin{proof} For $x(r)$ near $r=r_+$ we have $2\kappa_+ x(r)=\ln (r_+-r)+G(r),$ where $G$ is holomorphic near $r=r_+.$ Then $$w:=e^{2\kappa_+ x}=(r_+-r) e^{G(r)}. $$ We apply  the inverse function theorem to solve for $r$ as a function of $w$ near zero.  

Together with the similar analysis near
 $r=r_-$ it implies that there exists a constant $X_0>0$ such that for $\pm x >X_0,$ we have $r=r_\pm \mp F_\pm (e^{ 2\kappa_\pm x}),$ where $F_\pm (w)$ are real analytic on $[0, e^{\pm 2\kappa_\pm X_0})$
and holomorphic in the discs $\{ |w| < e^{\pm 2\kappa_\pm X_0}\}\in\C.$ 

Thus $r(x)$ has a holomorphic extension to the region $\{ z\in\C:\,\,|\Re x|>X_0\}.$ Since $r(x)$ is real analytic in $\R,$ one can extend it holomorphically to a region $\{ z\in\C:\,\,|\Re x|<X_0,\,\, |\Im z| <\epsilon\}$ for some $\epsilon >0.$ Unique continuation gives a holomorphic extension of $r(x)$ to a conic neighbourhood of the real axis. The bounds follows as in \cite{BachelotMotetBachelot1993}  (see also Lemma 2.1 in \cite{DaudeNicoleau2011}).\end{proof}

We consider the semiclassical Schr\"odinger operator 
\[\lb{semiP} P_h=(hD_x)^2+V_h(x),\qq V_h(x)=(\a(x))^2+h\a'(x),\qq x\in\R,
\]
where $\alpha>0$ is defined in (\ref{a2}). It follows from Proposition \ref{Prop4.1} that
the potential  $V_h(x)$ extends  to a holomorphic function in  a conic neighbourhood of the real axis given by $|\arg z|<\theta$ and satisfies there
$$|V_h(z)|\leq C\exp(-|z|/C),\qq\Re z\rightarrow\pm\infty.$$  Using the method of complex scaling \cite{SjostrandZworski1991} we can construct  meromorphic continuation of the resolvent $(P_h-E)^{-1}:\,\,C_0^\infty(\R)\rightarrow C^\infty (\R),$ $1<\arg E <2\pi,$ 
through the continuous spectrum $\arg E=0,$  to the unphysical sheet for $\arg E > -\theta.$ 
The poles of the extended resolvent are called resonance {\em energies}. Here, the Riemann surface of the resolvent $(P_h-E)^{-1}$  is the Riemann surface of the function $\sqrt{E}.$ Relation with the Schr\"odinger resonances considered in the previous section is given by $E=\lambda^2.$

\begin{figure}[htbp]
\includegraphics[width=\textwidth]{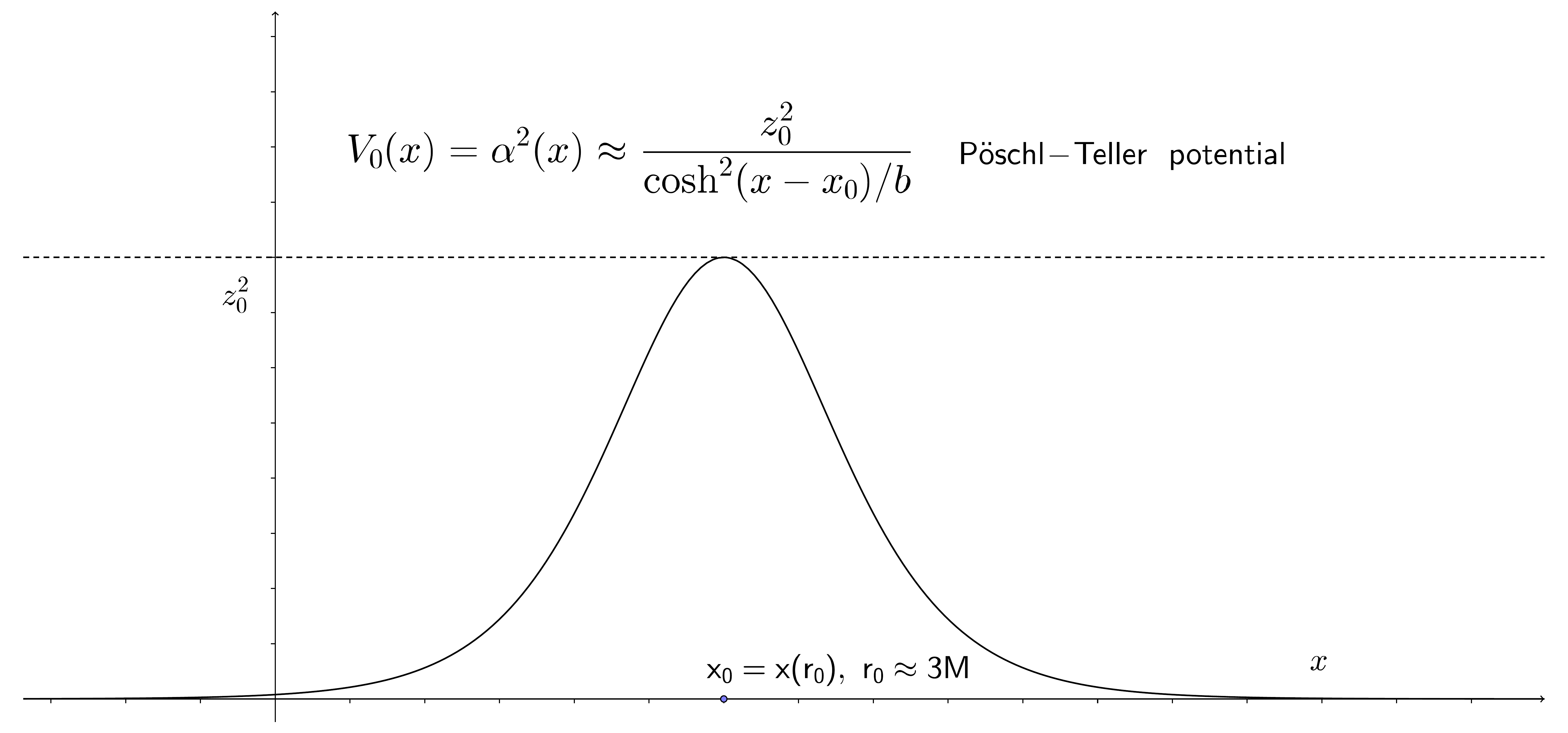}
 \caption{The potential $V_0.$ \label{Fig1} }
\end{figure}

Now, recall that   the principal symbol  of the potential $V_0(x)=\a^2(x)$ has unique non-degenerate maximum at $x_0,$ see Figure \ref{Fig1} (the profile of the potential is close to the P\"oshle-Teller potential, which is often used in the numerical calculations, see  \cite{Jing2004}).  From the dynamical point of view this means that the flow of the Hamiltonian given by the principal symbol of $P_h,$ $p(x,\xi)=\xi^2+V_0(x),$ has an unstable equilibrium point at $(x_0,0)$ --- a trapping point.  This is a very special case of the trapping sets which are closed hyperbolic trajectories and it is well-known  (see   \cite{GerardSjostrand1987}, \cite{Gerard1988} 
) that the associated resonances are close to the lattice of pseudopoles. The trapping point resonances --- resonances associated to the non-degenerate  critical point of the Hamiltonian  were considered in
 \cite{Sjostrand1986}. For one-dimensional operator $P_h$ these results imply that the resonances
 associated to the non-degenerate maximum  of the potential, barrier top resonances, are close to the string of pseudopoles parallel to the imaginary axis in $\C_-.$

The idea is to approximate the potential by its quadratic part near the maximum and to control the behaviour at infinity  by  the method of complex scaling. 
Hence, we consider the 
\begin{align}\lb{osc} 
&P^0(x,hD_x;h)=(hD_x)^2+z_0^2-\omega^2(x-x_0)^2,\\
\notag &z_0^2=V_0(x_0);\qq \omega^2=-\frac12 V_0''(x_0),
\end{align}
to which we can apply complex scaling formally:
$$P^0_\theta(x,hD_x;h)=P^0(w,hD_w;h)_{|\Gamma_\theta},\qq \Gamma_\theta=x_0+e^\theta \R\in\C, $$ so that with the coordinate $w$ on $\Gamma_\theta,$ $w=x_0+e^\theta y,$
$$P^0_\theta(y,hD_y;h)=e^{-2\theta}(hD_y)^2+z_0^2-e^{2\theta}\omega^2y^2.$$
Setting $\theta=i\pi/4$ we effectively turn our operator into multiple of the harmonic oscillator  
$$P^0_{\pi/4}(y,hD_y;h)=-i\left((hD_y)^2+\omega^2y^2\right)+z_0^2. $$ Since the eigenvalues of the harmonic oscillator $(hD_y)^2+\omega^2y^2$ are given by $\omega h(2k+1),$ $k=0,1,\ldots,$ we see that the eigenvalues of $P^0_{\pi/4}(y,hD_y;h)$ in the rectangle around $z_0^2=V_0(x_0)$ are given by the elements of the set of pseudopoles $$\G^0(h)=\left\{V_0(x_0)-ih\omega(2k+1);\,\,k=0,1,2,\ldots\right\},\qq \omega=\left(\frac12|V_0''(x_0)|\right)^\frac12,$$  which serves as an approximation modulo $o(h),$ $h\rightarrow 0,$ of barrier top resonances for $P_h.$

In this paper we apply a more refined construction. By conjugating the semiclassical operator $P_h$ with a Fourier integral operator microlocally near  the trapping point $(x_0,0)$ we can transform  $P_h$  into the  quantum  Birkhoff normal form (qBnf) \[\lb{Bi0}  z_0^2+\sum_{j=0}^\infty h^j f_j((hD_x)^2-\omega^2(x-x_0)^2),\quad f_0(\imath)=\imath+{\mathcal O}(\imath^2),\] 
%the $\equiv$ means to infinite order at $(x,\xi)=(x_0,0)$  (see \cite{IantchenkoSjostrand2002}). 
so that the approximation by $P^0$ as above is the zero order approximation. Here the Taylor expansions of $f_j$ at $0$ can be calculated iteratively. 

 The qBnf reduction in our context originates from the construction in \cite{Sjostrand1992} which was later applied to the trapping point resonances in \cite{KaidiKerdelhue2000} and extended to the resonances associated to a closed hyperbolic trajectory in \cite{Iantchenko2007}. In the later work we used a Birkhoff normal form construction for a quantum monodromy operator which is a Fourier integral operator associated to the non-linear Poincar{\'e} map along the closed trajectory (see \cite{Iantchenko1998} and \cite{IantchenkoSjostrand2002}).

The method of qBnf was successfully applied to the inverse semiclassical problems (see \cite{Guillemin1996}, \cite{Zelditch2002},
\cite{IantchenkoSjostrandZworski2002}) where the problem of reconstruction of the qBnf from the spectrum was studied. In \cite{Zworski2007} it was indicated how the
inverse spectral results based on wave invariants translates to
inverse results for resonances (see also \cite{Zelditch2002}).  In  \cite{Iantchenko2008}
 the inverse problem for the trapping point resonances was studied.

Now, we recall the construction of \cite{KaidiKerdelhue2000} and \cite{Iantchenko2008}.
Due to Proposition \ref{Prop4.1} the Schr{\"o}dinger operator $P_h$ defined in  (\ref{semiP}) has analytic potential which extends to a holomorphic function in a conic \neigh{} of the real axis so that hypothesis in \cite{KaidiKerdelhue2000} are satisfied. Then it can be 
transformed into the quantum  Birkhoff normal form $P^\infty$
\begin{equation}\label{Bnf} z_0^2+P^\infty\equiv U^*P_hU.
\end{equation}
Here $z_0^2=V_0(x_0)$ is the maximum of the leading term of the potential (for $h=0$),
 $U$ is analytic unitary \fourior{} microlocally defined near
$(0,0)$ and  $P^\infty$ is \pseudor{} with the symbol
\begin{equation}\label{F} F\sim\sum_{j=0}^\infty
h^jF_j(\imath ), \quad \imath=2\Omega=\xi^2-x^2,
\end{equation}
 with analytic $F_j,$  where the  principal and the sub-principal symbols are  given by
\begin{equation}\label{F0}
F_0(\imath)=\omega \imath +\frac12\omega b_{0,2}\,\imath^2
+{\mathcal O}(|\imath|^3),\quad F_1(\imath)=\omega b_{1,2}\, \imath+{\mathcal O}(|\imath|^2),
\end{equation}
coefficient $\omega$ is as in (\ref{osc}) and comes from the change of variables as explained in the next section, Eq.(\ref{lintr}),  $b_{0,2}, b_{1,2}$ are some numbers  calculated explicitly in the next section.

The equivalence relation $\equiv$ means  to infinite order for the symbols at
$(0,0)$ modulo ${\mathcal O}(h^\infty)$ (see \cite{IantchenkoSjostrand2002}).

Now, following  \cite{KaidiKerdelhue2000} we can apply the Helffer-Sj{\"o}strand theory  (see \cite{HelfferSjostrand1986})
and realize $P_h$ as acting in $H(\Lambda)$-spaces, where
$\Lambda\subset\cz^{2}$ is an IR-manifold which coincides with
$T^*(e^{i\pi/4}\rz)$ near $(0,0)$ and has the
property that $\forall\epsilon >0,\,\,\exists\delta >0$ such that
$(x,\xi)\in\Lambda,\,\,{\rm dist}\,((x,\xi),(0,0)) >\epsilon$
implies $|p(x,\xi) -z_0^2| >\delta.$

Then resonance energies can essentially (modulo an argument using a Grushin
reduction) be viewed as an eigenvalue problem for $P_h$ after the
complex scaling $ x=e^{i\pi/4} \tilde{x},$ $\tilde{x}\in\rz.$

By complex scaling of (\ref{Bnf}) one can show that the resonance energies
are close to the eigenvalues of
the  quantum Birkhoff normal form with symbol $z_0^2+\tilde{F}$ after  the complex
scaling $ x=e^{i\pi/4} \tilde{x},$ $\tilde{x}\in\rz.$
Here $$\tilde{F} \sim \sum_{j=0}^\infty
h^j\tilde{F}_j(\tilde{\imath}),\qq \tilde{F}_j(\tilde{\imath})=
F_j\left(\frac{1}{i}\tilde{\imath}\right),$$
 $F$  is as in (\ref{F}) and
$\frac{1}{i}\tilde{\imath}=\frac{1}{i}(\tilde{\xi}^2+\tilde{x}^2)=\xi^2-x^2,$
$\xi=e^{-i\pi/4}\tilde{\xi},$ $x=e^{i\pi/4}\tilde{x}.$ 
The result of Kaidi and  Kerdelhue  \cite{KaidiKerdelhue2000} applied to operator $P_h$ as in (\ref{semiP}) in one dimensional case states
 that\\
{\em The resonance energies $E_k=\lambda_k^2$ of $P_h$ in rectangle $]z_0^2-\epsilon_0,z_0^2+\epsilon_0[
-i[0,h^{\delta}],$ $\delta>0,$ are simple labeled by $k\in\nz$ and of the form ${\displaystyle z_0^2+\sum_{j=0}^\infty h^j\tilde{F}_j((2k+1)h)}.$}

Now, using (\ref{F0}) we get
$$\tilde{F}_0(\imath)= - i\omega\,\imath -\frac12\omega b_{0,2}\,\imath^2
+\mathcal{O}(|\imath |^3),\quad \tilde{F}_1(\imath)=-i\omega b_{1,2}\, \imath+{\mathcal O}(|\imath|^2),$$ and 
in the leading order as $h\rightarrow 0$  the resonance energies are given by
\begin{align}
E_k&=z_0^2- ih\omega (2k+1)-\frac12\omega b_{0,2}h^2(2k+1)^2-h^2i\omega b_{1,2} (2k+1)+{\mathcal O}(h^3)\nonumber\\
&=z_0^2- ih\omega (2k+1)\left(1+h\left[\frac{1}{2i} b_{0,2}(2k+1)+b_{1,2} \right]\right)+{\mathcal O}(h^3.)\lb{pseudopoles}
\end{align}

Note that, applying the semiclassical inverse results from \cite{IantchenkoSjostrandZworski2002} to the 
resonances
(see \cite{Zworski2007} and \cite{Zelditch2002}) we know that the full qBnf  can be reconstructed from the resonances.
% Together with Theorem \ref{th_CdV_Gui} this implies Theorem \ref{th-resinv}

An interesting question would be if one can reconstruct the (Taylor series of) the potential from the resonances. In the case of operator $P_0$ (for $h=0$ in (\ref{semiP})) the answer is positive if $V_0'''(x_0)\neq 0,$ which follows from   \cite{CdVerdiereGuillemin2011} and \cite{Iantchenko2008}. In the next section we apply the method of  \cite{CdVerdiereGuillemin2011} to $P_h$ and calculate $b_{0,2}$ and $b_{1,2}.$

\subsection{Explicit calculation of qBnf}\lb{ss-Exp}
Here we apply the method from \cite{CdVerdiereGuillemin2011}, Section 8, in order to calculate few leading coefficients in qBnf. 

Recall that the leading term of the potential in (\ref{Schr}) $V_0(x)=\a^2(x)$ has a non-degenerate maximum at $x_0=x(r_0),$ where
$$
 r_0=\frac{3M+\sqrt{(3M)^2-8Q^2}}{2}$$ and  ${V_0(x_0)=r_0^{-4}\left(Mr_0-Q^2-\frac{\Lambda}{3}r_0^4,\right)}$ is the maximum value of $V_0.$ We will use the following notations: 
$$z_0^2=V_0(x_0),\quad \omega=\sqrt{\frac12|V_0''(x_0)|},\quad 2^\frac12\omega=\sqrt{|V_0''(x_0)|},\quad V_0''(x_0)=-2\omega^2.$$ Using 
$$\left(\frac{dx}{dr}\right)^{-2}=F^2(r),\qq V_0=\frac{F(r_0)}{r_0^2},\qq \left.\frac{d^2V_0}{dr^2}\right|_{r=r_0}=\frac{2}{r_0^6}\left(-3Mr_0+4Q^2\right)$$ 
we get (\ref{somef}),
\begin{align*}
V_0''(x_0)&=\left(\frac{dx}{dr}\right)^{-2}\left.\frac{d^2V_0}{dr^2}\right|_{r=r_0}\\
&=-2\left(\frac{3M}{r_0}-\frac{4Q^2}{r_0^2}\right)V_0^2(x_0)=-2\left(1-\frac{2Q^2}{r_0^2}\right)V_0^2(r_0)
\end{align*}
and $$V_0'''(x_0)=\frac{4}{r_0}\left(11Mr_0-18Q^2-8Mr_0^3+12Q^2r_0^2+\frac43\Lambda\left[r_0^4-r_0^6\right]\right)V_0^3(x_0).$$

% We have also \begin{align*}
%V_0''(x_0)=&\left(\frac{4Q^2}{r_0^2}-2\right)V_0^2(r_0),\qq \frac{3M}{r_0}-\frac{4Q^2}{r_0^2}= 1-\frac{2Q^2}{r_0^2},\\
%V_0'''(x_0)=&\frac{4}{r_0}\left(11Mr_0-18Q^2-8Mr_0^3+12Q^2r_0^2+\frac43\Lambda\left[r_0^4-r_0^6\right]\right)V_0^3(x_0).\end{align*}

Now, the Taylor expansion of $V_h$ is given by
\begin{align*}
V_h(x)&=z_0^2-\omega^2(x-x_0)^2+\frac{V_0(x_0)'''}{6}(x-x_0)^3\\
&\quad +\frac{V_0(x_0)''''}{24}(x-x_0)^4+{\mathcal O}(x-x_0)^5\\
&\quad + h\bigg[-\frac{1}{z_0}\omega^2(x-x_0)\\
&\qquad\quad +\frac12\left(\frac12 z_0^{-3}\omega^2+z_0^{-1}V_0'''(x_0)\right)(x-x_0)^2+{\mathcal O}(x-x_0)^3\bigg].
\end{align*}

Symbol of the operator $P_h$ (see (\ref{semiP})) is given by
$$p_h(x,\xi)=\xi^2+V_h(x).$$ We can put $x_0=0.$ Then perform the linear symplectic transformation
\[\lb{lintr}x\mapsto\omega^\frac12x,\qq\xi\mapsto\omega^{-\frac12}\xi.\] The transformed symbol is denoted 
by the same letter
\begin{align*}p_h(x,\xi)&=z_0^2+\omega(\xi^2-x^2)+\frac{V_0'''(x_0)}{6\omega^\frac32}x^3+\frac{V_0''''(x_0)}{24\omega^2}x^4+{\mathcal O}(x^5)\\
&\quad +h\left[-\frac{\omega^\frac32}{z_0}x+\frac12\left(\frac12 z_0^{-3}\omega+\frac{1}{z_0\omega}V_0'''(x_0)\right)x^2+{\mathcal O}(x^3)\right] .\end{align*}
We will work with the Hamiltonian
\begin{align*}H:=\frac{1}{2\omega}p_h(x,\xi)&=\frac{z_0^2}{2\omega}+\frac12(\xi^2-x^2)+\frac{V_0'''(x_0)}{12\omega^\frac52}x^3+\frac{V_0''''(x_0)}{48\omega^3}x^4+{\mathcal O}(x^5)\\
&\quad + h\left[-\frac{\omega^\frac12}{2z_0}x+\frac14\left(\frac12 z_0^{-3}+\frac{1}{z_0\omega^2}V_0'''(x_0)\right)x^2+{\mathcal O}(x^3)\right] .\end{align*}

Following the notations in \cite{CdVerdiereGuillemin2011} we  put
$$E_0:=\frac{z_0^2}{2\omega},\qq\Omega=\Omega_-:=\frac12(\xi^2-x^2),\qq a_3:=\frac{V_0'''(x_0)}{12\omega^\frac52},\qq a_4=\frac{V_0''''(x_0)}{48\omega^3}.$$ In addition, we denote
$$ c_1:=- \frac{\omega^\frac12}{2z_0},\qq c_2:=\frac14\left(\frac12 z_0^{-3}+\frac{1}{z_0\omega^2}V_0'''(x_0)\right).$$ So we consider the classical symbol
%$$H=E_0+\Omega+a_3x^3+{\mathcal O}(x^4)+h\left(c_1^\pm x+ c_2^\pm x^2 +{\mathcal O}(x^3)\right)$$ or
\[\lb{Hterm}H=E_0+\Omega+a_3x^3+\sum_{j=4}^\infty a_j x^j +h\left(c_1 x+ c_2 x^2 +\sum_{j=3}^\infty c_j x^j\right), \] which we want to transform into 
the quantum Birkhoff normal form 
\begin{align}\lb{qBnf}
H^{\rm qBnf}&=E_0+\Omega+b_{0,2}\Omega^2+\sum_{j=3}^\infty b_{0,j}\Omega^j\\
\notag &\quad +h\left(b_{1,2} \Omega+ b_{1,4} \Omega^2 +\sum_{j=3}^\infty b_{1,2j}\Omega^j\right)+\cdots.
\end{align}
The difference from the situation considered in  \cite{CdVerdiereGuillemin2011} is that the symbol $H$ in (\ref{Hterm}) also contains the $h$-dependent terms. This leads to a modification of the algorithm of \cite{CdVerdiereGuillemin2011}  and results in the presence in the Birkhoff normal form $H^{\rm Bnf}$ also the odd powers of $h,$ whereas in  \cite{CdVerdiereGuillemin2011} only even powers of $h$ were present (see formula (1) there).  As a consequence,  the method of \cite{CdVerdiereGuillemin2011} does not allow us to reconstruct the potential from the qBnf coefficients. 
 
Following Section 3 in \cite{CdVerdiereGuillemin2011}  we introduce the product rule of symbols $a(x,\xi),$ $b(x,\xi)$ of the Weyl quantized \pseudor s (the Moyal product) as follows:
$$a\star b:= \sum_{j=0}^\infty\frac{1}{j!}\left(\frac{h}{2i}\right)^j\{a,b\}_j$$ with
$$\{a,b\}_j:=\sum_{n=0}^j\ma j\\n\am (-1)^n\partial_x^n\partial_\xi^{j-n}a\partial_x^{j-n}\partial_\xi^nb. $$ We will also use the Moyal bracket
$$[a,b]^\star:=a\star b-b\star a. $$ Note that
$$\frac{i}{h}[a,b]^\star=\sum_{j=0}^\infty\frac{1}{2j+1}\left(\frac{h}{2i}\right)^{2i}\{a,b\}_{2j+1}.$$
In order to reduce $H$ to the qBnf, we perform transformation $$H\mapsto H^{\rm qBnf}=H_S=e^{iS/h} \star H\star  e^{-iS/h} =\exp{\left(\frac{i}{h}{\rm  ad}\,(S)^\star\right)} H,$$ where
$$S=S_0+hS_1+h^2S_2+\cdots,\qq S_0=S_0^3+S_0^4+\cdots,\qq S_0^3=\sum_{i+j=3}s_{i,j}x^i\xi^j,$$ and
$$S_1=S_1^1+S_1^2+\cdots,\quad S_1^1=s_{1,0}x+s_{0,1}\xi,\ldots $$
Here
\begin{align}\lb{ad}
\exp{\left(\frac{i}{h}{\rm  ad}\,(S)^\star\right)} H&=\exp{\left(\frac{i}{h}[S, .]^\star\right)} H\\
\notag &=H+\frac{i}{h}[S,H]^\star +\frac12 \left(\frac{i}{h}\right)^2[S,[S,H]^\star]^\star+\cdots\end{align}
is a convergent formal power series in $x^k\xi^nh^m$,
and
\begin{align*}
&\frac{i}{h}[S,H]^\star=\{S,H\}_1-\frac{1}{24}h^2\{S,H\}_3+\cdots,\quad \{S,H\}_1=S'_\xi H_x' -S'_xH'_\xi,\\
&\{S,H\}_3=S^{(3)}_{\xi\xi\xi} H^{(3)}_{xxx}-3S^{(3)}_{\xi\xi x} H^{(3)}_{xx\xi}+3 S^{(3)}_{\xi x x} H^{(3)}_{x\xi\xi}-S^{(3)}_{xxx} H^{(3)}_{\xi\xi\xi}.
\end{align*}

We want to reduce  (\ref{ad}) to (\ref{qBnf}). 
%Starting with $h$-independent terms, we identify the terms of order 3 and 4.
We choose $S_0^3, S_0^4,\ldots$ so that
\begin{align}
(x,\xi)^3:\qq& \quad\ a_3x^3+\{S_0^3,\Omega\}_1=0,\lb{no_h}\\
(x,\xi)^4:\qq& \quad\ a_4x^4+\{S_0^3,a_3x^3\}_1+\{S_0^4,\Omega\}_1\lb{first_h}\\
&\quad +\frac12\{S_0^3,\{S_0^3,\Omega\}_1\}_1
-\frac{1}{24} h^2\{S_0^3,a_3x^3\}_3\nonumber\\
&=b_{0,2}\Omega^2+h^2b_{2,0}+\cdots\nonumber
\end{align}
as $$\frac{i}{h}[S_0,H]^\star=\{S_0,H\}_1-\frac{1}{24} h^2\{S_0,H\}_3.$$

Then  from (\ref{no_h}) it follows that ${\displaystyle S_0^3=a_3x^2\xi-\frac23a_3\xi^3
%,\,\, b_{2,0}=a_3^2
.}$ Note the difference of sign with formula (2) in \cite{CdVerdiereGuillemin2011} in front of the first term above (see remark after (\ref{b02}) below).
Using (\ref{no_h}) equation (\ref{first_h}) splits into $2$ equations
\begin{align}
& a_4x^4+\frac12\{S_0^3,a_3x^3\}_1+\{S_0^4,\Omega\}_1=b_{0,2}\Omega^2,\lb{secab}\\
& -\frac{1}{24} h^2\{S_0^3,a_3x^3\}_3
=h^2b_{2,0}+\cdots.\lb{hdep}
\end{align}
We emphasize that contrary to  \cite{CdVerdiereGuillemin2011}  equation (\ref{hdep}) contains more terms of order ${\mathcal O}(h^2),$ which will be specified later.

The coefficient $b_{0,2}$ is obtained from  (\ref{secab}) where $S_0^4$ is obtained from the  equation for $ (x,\xi)^5$ with zero right hand side.  The $h-$independent reduction is done exactly as in  \cite{CdVerdiereGuillemin2011}, and it follows from  Theorem 8.1 there   that
\[\lb{b02} b_{0,2}=\frac{15}{4}a_3^2+\frac32 a_4.
\] Note the difference of sign with \cite{CdVerdiereGuillemin2011} in front of the first term in (\ref{b02}) as we consider $\Omega=\Omega_-=\frac12(\xi^2-x^2)$ instead of $ \Omega_+=\frac12(\xi^2+x^2).$

After all $h$-independent terms are reduced to the Bnf we arrive at 
\begin{align*}
H_{S_0}&:=e^{iS_0/h} \star H\star  e^{-iS_0/h}\\
&=E_0+\Omega+b_{0,2}\Omega^2+\sum_{j=3}^\infty b_{0,2j}\Omega^j\\
&\quad+h\left(c_1 x+ c_2 x^2 +\{S_0^3,c_1x\}_1+{\mathcal O} (x,\xi)^3\right) +{\mathcal O} (h^2).
\end{align*}
We will keep in mind the following terms
$$c_1 x+ c_2 x^2 +\{S_0^3,c_1x\}_1=c_1 x+ (c_2+c_1a_3) x^2-c_1 a_32\xi^2.$$ 
Now,
\begin{align*}
H_{S_1}&:=e^{ihS_1/h} \star H_{S_0}\star  e^{-ihS_1/h}\\
&=\exp{\left(\frac{i}{h}[hS_1, .]^\star\right)} H_{S_0}=H_{S_0}+\frac{i}{h}[hS_1,H_{S_0}]^\star \\
&\quad +\frac12 \left(\frac{i}{h}\right)^2[hS_1,[hS_1,H_{S_0}]^\star]^\star+\cdots,\\
\frac{i}{h}[&hS_1,H_{S_0}]^\star=\{hS_1,H_{S_0}\}_1-\frac{1}{24}h^2\{hS_1,H_{S_0}\}_3+\cdots.
\end{align*}
We choose $S_1^1, S_1^2,\ldots$ so that
\begin{align*}
h(x,\xi)^1:\qq& hc_1x+\{hS_1^1,\Omega\}_1=0,\\
h(x,\xi)^2:\qq& h(c_2+c_1a_3)x^2-hc_1a_32\xi^2+h\{S_1^2,\Omega\}_1\\
&+\frac12\{hS_1^1,\{hS_1^1,\Omega\}_1\}_1=hb_{1,2}\Omega.
\end{align*}
The first equation implies that $S_1^1=c_1\xi,$ then from the second one we get
\[\lb{b12}b_{1,2}=-3c_1a_3-c_2\] and 
 $$\frac12\{hS_1^1,\{hS_1^1,\Omega\}_1\}_1=-\frac12\{hS_1^1,hc_1 x\}_1=-\frac12h^2c_1^2.$$ Combining this term with  another $h^2{\mathcal O}(1)$ term appeared in (\ref{first_h}) and (\ref{hdep}) we get equation
$$h^2{\mathcal O}(1):\  -\frac{1}{24} \{S_0^3,a_3x^3\}_3-\frac12 c_1^2=b_{2,0},$$ where
  $S_0^3=a_3x^2\xi-\frac23a_3\xi^3.$ Then we get
\[\lb{b20}a_3^2-\frac12 c_1^2=b_{2,0}. \] Note that if $c_1=0$ we recover the second formula in Theorem 8.1 from \cite{CdVerdiereGuillemin2011}.

Following this strategy we can reconstruct all the coefficients of the  qBnf which can be proved by induction as in Section 9 in \cite{CdVerdiereGuillemin2011}
\begin{align*}
H^{\rm Bnf}&=E_0+\Omega+b_{0,2}\Omega^2+\sum_{j=3}^\infty b_{0,j}\Omega^j\\
&\quad +h\left(b_{1,2} \Omega+ b_{1,4} \Omega^2 +\sum_{j=3}^\infty b_{1,2j}\Omega^j\right)+\cdots. 
\end{align*}

\section{Application to \dSRN resonances}\lb{s-dSRN}
Recall that  (\ref{DiracSystem}) is written in semiclassical way as follows
$$
\cD\psi\equiv\left[ h\s_3D_x-\alpha (x)\s_1\right]\psi=z \psi,\,\, z=\lambda/n=\lambda h,\qq n=(l+1/2),$$ 
with the ``Planck constant'' $ h=1/n$ and exponentially decreasing potential satisfying
(\ref{expdec}).
%We consider also the Dirac operator with "plus" sign  $\cD_{\alpha}$.
Recall that operation $\{\cdot\}^{\rm S}\in\C_-$ denotes the mirror reflection of the set $\{\cdot\}\in\C_-$ with respect to $i\R$ (see (\ref{mirror})). 

 Then  (see Lemma \ref{L-UnD}) 
${\displaystyle {\rm Res}\,(P)\setminus\{0\}={\rm Res}\,(\cD)\cup{\rm Res}^{\rm S}\,(\cD)\subset\C_-},$ where ${\displaystyle {\rm Res}\,(P)}$ is the resonance set  for the Schr{\"o}dinger operator $P,$
\[\lb{semiP2}P=h^2(D_x)^2+V_h(x),\qq V_h(x)=(\a(x))^2+h\a'(x).\]

Thanks to the exponential decrease of the potential at both infinities (\ref{expdec}) we have  the following result  on the resonance free domain for the operator  $P_n:= D_x^2+ n^2\alpha^2(x)+n\alpha'(x), $  which follows from \cite{SaBarretoZworski1997} proved there for the operator $D_x^2+ n^2\alpha^2(x).$ 
\begin{proposition}\lb{th-resfree} For $R$ large enough, operator $P_n=D_x^2+ n^2\alpha^2(x)+n\alpha'(x) $ has no resonance 
 in $[R,n/R]+i[-C_0,0].$ Here $n=l+1/2.$
\end{proposition}
A detailed presentation of the techniques needed in the proof is available in Section 5.2 of \cite{Datchev2010} and Section 5 of  \cite{Datchev2012}.

%The techniques needed in \cite{SaBarretoZworski1997} to prove this proposition is presented  in Section 5.2 of \cite{Datchev2010} and Section 5 of  \cite{Datchev2012}.

In the previous section we approximated the resonance energies $E=\lambda^2$  for $P$ defined in  (\ref{semiP2}) by pseudopoles ({\ref{pseudopoles}). 
If $\lambda_k$ are the  resonances for $P$, then the corresponding resonance energies $\lambda_k^2h^2$  in rectangle $]z_0^2-\epsilon_0,z_0^2+\epsilon_0[
-i[0,h^{\delta}]$ are simple labeled by $k\in\nz$ and of the form
$$\mu_{k}^2h^2=z_0^2+\sum_{j=0}^\infty h^j\tilde{F}_j((2k+1)h),$$ $$ \tilde{F}_0(\imath)= - i\omega\,\imath -\frac12\omega b_{0,2}\,\imath^2
+\mathcal{O}(|\imath |^3),\quad \tilde{F}_1(\imath)=-i\omega b_{1,2}\, \imath+{\mathcal O}(|\imath|^2).$$

Now,  the resonances $ \lambda_k$ for $P$  in rectangle \[\lb{rect}(l+1/2)\cdot]z_0-\epsilon_0',z_0+\epsilon_0'[
-i[0,(l+1/2)^{1-\delta/2}]\] are simple labeled by $k\in\nz$ and of the form $$\mu_{k}=(l+1/2)\left(z_0^2+\sum_{j=0}^\infty (l+1/2)^{-j}\tilde{F}_j\left(\frac{2k+1}{l+1/2}\right)\right)^{1/2}. $$

In order to get the leading terms in the expansion of resonances in rectangle
(\ref{rect}) we use (\ref{pseudopoles}): if $ \lambda$ is a resonance for $P$  then $\lambda^2h^2$ is approximated by
 $$\mu_{k}^2h^2=E_k=z_0^2- ih\omega (2k+1)\left(1+h\left[\frac{1}{2i} b_{0,2}(2k+1)+b_{1,2} \right]\right)+{\mathcal O}(h^3)$$
 which implies 
\begin{align*}
 \mu_{k}&= z_0h^{-1}- \frac12z_0^{-1}i\omega (2k+1)\\
 &\quad\times\left(1+h\left[-\frac{1}{4iz_0^2}\omega(2k+1)+\frac{1}{2i} b_{0,2}(2k+1)+b_{1,2} \right]\right)\\
 &\quad+{\mathcal O}(h^2).
\end{align*}
Here $z_0=\alpha(x_0)=\sqrt{V_0(x_0)}$ is the maximum value of $\alpha$ and
$\omega=\sqrt{\frac12|V_0''(x_0)|}.$ Now, as $h=(l+1/2)^{-1},$ we get
\begin{align*}
\mu_{k}&=z_0(l+1/2)- \frac{i\omega}{2 z_0}  (2k+1)\\
&\quad -\frac{i\omega}{2 z_0} \frac{(2k+1)}{(l+1/2)}\left[-\frac{1}{4iz_0^2}\omega(2k+1)+\frac{1}{2i} b_{0,2}(2k+1)+b_{1,2} \right]\\
&\quad +{\mathcal O}((l+1/2)^{-2}). \end{align*}

Together with Proposition \ref{th-resfree} and Lemma \ref{L-UnD}  we get Theorem \ref{th-dSRN}.

\section{Application to \dSS resonances}\lb{s-dSS}
Here we show how the same method works in the \dSS case  and prove the formulas given in the Remark 2 to Theorem \ref{th-dSRN}.
Recall that in the \dSRN case the governing equation was
$$\left(D_x^2+(l+1/2)^2\alpha^2+(l+1/2)\alpha'\right)\psi=\lambda^2\psi.$$

In the \dSS we consider the following equation instead (see \cite{SaBarretoZworski1997})
\[\lb{desit}\left(D_x^2+\alpha^2[l(l+1)+2\alpha \alpha' r^3+2\alpha^2 r^2]\right)\psi=\lambda^2\psi,\] where (putting $Q=0$)
$$\a^2(x)=\frac{F(r(x))}{r^2(x)},\qq F(r)=1-\frac{2M}{r}-\frac{\Lambda}{3}r^2.$$
 We put $h=(l(l+1))^{-1/2}.$ Then equation is transformed into the semiclassical one  
\begin{align}\lb{semi-dSS}
&P_h u:= \left(h^2D_x^2+W_h\right)u=Eu,\\
\notag &W_h=\alpha^2[1+h^2(2\alpha \alpha' r^3+2\alpha^2 r^2)],\quad E=h^2\lambda^2.
\end{align}
Note that the principal symbols in \dSS and \dSRN cases coincide  $W_0=V_0$  (after we put charge $Q=0$ in the later case).

As in \dSRN case we can calculate explicitly the coefficients of the qBnf as in Section \ref{ss-Exp}. Repeating the procedure we  consider the classical symbol
$$H=E_0+\Omega+a_3x^3+{\mathcal O}(x^4)+h^2\left(d_0+ d_1x+{\mathcal O}(x^2)\right),\qq d_0=\frac{V_0^2(x_0)}{\omega}9M^2,$$ 
 which we want to transform into 
the qBnf 
\begin{align*}
H^{\rm Bnf}&=E_0+\Omega+b_{0,2}\Omega^2\\
&\quad +\sum_{j=3}^\infty b_{0,j}\Omega^j+h^2\left(b_{2,0}+b_{2,2} \Omega+ b_{2,4} \Omega^2 +\sum_{j=3}^\infty b_{2,2j}\Omega^j\right)+\cdots. 
\end{align*} 
Note that symbol $H$ contains $h^2$ terms and not $h$ terms as it was in the \dSRN case. This leads to  the qBnf containing only even powers of $h,$ as it was in  \cite{CdVerdiereGuillemin2011}, formula (1).

In order to get coefficient $b_{0,2}$ we can use (\ref{first_h}) and (\ref{hdep}), where in the left hand side we add  the term $h^2d_0.$ Using $$S_0^3=a_3x^2\xi-\frac23 a_3\xi^3,\qq
-\frac{1}{24}h^2\{S_0^3,a_3x^3\}_3=a_3^2 h^2,$$ we get 
\[\lb{b20bis}b_{2,0}=a_3^2+d_0=\frac{(V_0'''(x_0))^2}{12^2\omega^5}+\frac{V_0^2(x_0)}{\omega}9M^2.\] Note that if $d_0=0$ we recover the second formula in Theorem 8.1 from \cite{CdVerdiereGuillemin2011}.
% together with (\ref{b02}) obtained as before.

 The eigenvalues of the complex-scaled qBnf in rectangle $]z_0^2-\epsilon_0,z_0^2+\epsilon_0[
-i[0,h^{\delta}]$ are simple labeled by $k\in\nz$ and of the form
$$z_0^2+\sum_{j=0}^\infty h^{2j}\tilde{F}_{2j}((2k+1)h)$$ where
$$\tilde{F}_0(\imath)= - i\omega\,\imath -\frac12\omega b_{0,2}\,\imath^2
+\mathcal{O}(|\imath |^3),\quad \tilde{F}_2(\imath)=2\omega b_{2,0}+{\mathcal O}(|\imath|).$$

In the leading order as $h\rightarrow 0$ we get
\begin{align}
E_k&=V_0(x_0)- ih\omega (2k+1)-\frac12\omega b_{0,2}h^2(2k+1)^2+h^22\omega b_{2,0}+{\mathcal O}(h^3)\nonumber\\
&=V_0(x_0)- ih\omega (2k+1)\left(1+h\frac{1}{2i} b_{0,2}(2k+1)\right)+h^22\omega b_{2,0}+{\mathcal O}(h^3).\lb{pseudopolesbis}
\end{align}

Now,  the resonance energies $ E=\lambda^2$ for the problem (\ref{desit}) in rectangle  \[\lb{rectbis}(l(l+1))\cdot]z_0^2-\epsilon_0,z_0^2+\epsilon_0[
-i[0,[l(l+1)]^{1-\delta/2}]\] are simple labeled by $k\in\nz$ and of the form $$\mu_{k}^2=l(l+1)\left(z_0^2+\sum_{j=0}^\infty [l(l+1/2)]^{-j}\tilde{F}_{2j}\left(\frac{2k+1}{[l(l+1)]^{1/2}}\right)\right). $$

In order to get the leading terms in the expansion of resonances $\lambda=\sqrt{E}$ in rectangle

 \[\lb{rectbisbis}[l(l+1))]^\frac12 \cdot]z_0-\epsilon_0,z_0+\epsilon_0[
-i[0,[l(l+1)]^{1/2-\delta/4}]\]
we use (\ref{pseudopolesbis}) 
 $$h^2\mu_{k}^2=E_k=z_0^2- ih\omega (2k+1)+h^2\left(\frac{\omega}{2} b_{0,2}(2k+1)^2-2\omega b_{2,0}\right)+{\mathcal O}(h^3).$$
 
Here $z_0^2=\alpha^2(x_0)=V_0(x_0)$ is the maximum value of $V_0$ and
$\omega=\linebreak \sqrt{\frac12|V_0''(x_0)|}.$ Now, as $h=[l(l+1)]^{-1/2}$ we get
\begin{align*}
\mu_{k}&=z_0[l(l+1)]^{\frac12}- \frac{i\omega}{2z_0} (2k+1)\\
&\quad -[l(l+1)]^{-\frac12}\frac{\omega}{2z_0} \left(\frac{1}{2} b_{0,2}(2k+1)^2-2 b_{2,0}\right)
+{\mathcal O}([l(l+1)]^{-1}). \end{align*}

Combining with Proposition \ref{th-resfree} and using that $[l(l+1)]^{\frac12}= (l+1/2)-\frac18(l+1/2)^{-1}+{\mathcal O}((l+1/2)^{-3})$ we get the result. 
\begin{align*}
\mu_{k,l}&=\omega^\frac12\left( (l+l/2) -i(k+1/2)\right)\\
&\quad -\omega^\frac12(l+1/2)^{-1}\left(\frac18+\frac{1}{4} b_{0,2}(2k+1)^2- b_{2,0}\right)+{\mathcal O}\left(l+1/2\right)^{-2}.\end{align*}

\end{document}